# Thermal expansion coefficients of high thermal conducting BAs and BP materials

Sheng Li[1], Keith M. Taddei[2], Xiqu Wang[3], Hanlin Wu[1], Jörg Neuefeind[2], Davis Zackaria[1], Xiaoyuan Liu[1], Clarina Dela Cruz[2], Bing Lv[1,a)]

[1] Department of Physics, The University of Texas at Dallas, Richardson, TX 75080, USA

[2] Neutron Scattering Directorate, Oak Ridge National Laboratory, Oak Ridge, Tennessee, 37831, USA

[3] Department of Chemistry, University of Houston, Houston, TX 77204, USA

a) author to whom correspondence should be addressed: blv@utdallas.edu

**ABSTRACT**

Recent reported very high thermal conductivities in the cubic boron arsenide (BAs) and boron phosphide (BP) crystals could potentially provide a revolutionary solution in the thermal management of high power density devices. To fully facilitate such application, compatible coefficient of thermal expansion (CTE) between the heat spreader and device substrate, in order to minimize the thermal stress, need to be considered. Here we report our experimental CTE studies of BAs and BP in the temperature range from 100K to 1150K, through a combination of X-ray single crystal diffraction and neutron powder diffraction. We demonstrated the room temperature CTE, $3.6 \pm 0.15 \times 10^{-6}$/K for BAs and $3.2 \pm 0.2 \times 10^{-6}$/K for BP, are more compatible with most of the semiconductors including Si and GaAs, in comparison with diamond, and thus could be better candidates for the future heat spreader materials in power electronic devices.

The cubic boron arsenide (BAs) and boron phosphide (BP) materials have attracted considerable attention recently due to their reported very high thermal conductivity.[1-5] The measured room temperature thermal conductivities are not only much higher than that of commercial silicon and wide-band-gap semiconductors such as GaAs, GaN and β-$Ga_2O_3$, but are also much higher than that of most heat spreaders currently used in the high-power electronics such as SiC, Ag, and Cu. Therefore, these materials hold promise as heat spreaders to relieve the increasing thermal dissipation issues faced by the electronic and power electronics industry.

A thermal heat spreader material is used as passive cooling interface to extract heat from a source (hot spot) and spread it within a minimum amount of time over a large area of the heat sink surface. For a heat spreader to efficiently remove heat from electronic devices, two critical factors must be simultaneously considered: the thermal conductivity and the difference between the coefficients of thermal expansion (CTE) of the heat spreader and the sinking materials.[6-10] A large mismatch in CTE between the heat spreader and the sinking material will generate significant mechanical stress which is harmful to both the performance and mechanical stability of devices. Compatibility between the CTEs of heat spreaders and device substrates is fundamentally important for electronic devices and engineering constructions.

For the current state-of-the-art flip chip package, copper or copper composite is most commonly used as the heat spreader because of its high thermal conductivity (~400 W/m/K) at room temperature. However, copper has a much higher CTE (16.5 x $10^{-6}$ /K) than silicon (2.63 x $10^{-6}$ /K ) which unavoidably creates thermally induced stress. Alternatively, diamond is used as the heat spreader because diamond has the highest isotropic thermal conductivity (>2000 W/m/K). However, diamond is expensive, difficult to fabricate, and it has a way smaller CTE (0.8 x $10^{-6}$ /K) than silicon, not to mention that wafer-scale CVD diamond spreaders are limited by the availability of large size and uniform stress free thickness. In order to make a CTE compatible heat spreader, diamond/copper composites made of diamond particles in copper metal have been developed.[11-13] However, the desirable increase of CTE is accompanied by an undesirable significant reduction of the thermal conductivity. Bulk diamond composite with the binder metal or metal infiltration is therefore often used as board level heat sink rather than as a chip-level heat spreader.

Therefore, the outlook for using cubic BAs and BP as future heat spreaders in electronics and power devices is rather appealing. The theoretical calculated CTEs of BAs (3.04×$10^{-6}$ $K^{-1}$) and BP (2.8 ×$10^{-6}$ $K^{-1}$) are very close to that of Si (2.63×$10^{-6}$ $K^{-1}$), which could significantly reduce the thermal stress of power electronic materials – if they are growth compatible. The calculated CTEs of BAs and BP are also closer to the CTEs of WBG semiconductors such as GaN, β-$Ga_2O_3$ and SiC, compared with that of diamond. However, experimental information about the CTEs of BAs and BP is lacking at this moment. In addition, future device integration likely requires the growth of BAs or BP films directly on the semiconducting devices. Such heteroepitaxial structures might undergo several heating–growth–cooling cycles during the process. This indicates that not only the interface stress induced by lattice mismatch, but also the thermal stress caused during the heating-cooling process, needs to be considered. Precise knowledge regarding thermal mismatch and stress, not only at room temperature but also in the extended temperature range close to growth temperature, is therefore important. In this paper we report thermal expansion measurements of the high thermal conducting materials BAs and BP in the extended temperature range from 100K to 1150K, through a combination of X-ray single crystal diffraction and neutron powder diffraction. The obtained CTEs of BAs and BP, are slightly higher than the predicted value, yet have the closest CTE to that of Si and GaAs of suitable materials. Given the smaller lattice mismatch of BAs/BP with Si and GaAs, compared to diamond, it is possible to directly deposit BAs/BP to Si or GaAs as a heat spreader without needing solder joints or other thermal interface materials. This allows for further enhancement to the efficiency of heat dissipation, and therefore potentially provide a revolutionary solution in the thermal management of high power density devices.

Powder BAs and BP samples were made through solid state reaction with boron powder (Alfa Aesar 99.9%) and arsenic chunks (Alfa Aesar 99.999%) or red phosphorus pieces (Alfa Aesar 99.999%) at 800 °C (for BAs) or 900 °C (for BP) for 2 days. The samples were ground and heat treated for at least 3 times to achieve homogeneity. Excess amounts of arsenic or phosphorus were added to avoid vacancies of As or P. The excess arsenic or phosphorus deposits at the cold end of the container during the reaction and thus does not contaminate the samples. Single crystals of BAs were grown through chemical vapor transport method while crystals of BP were synthesized using metal flux method described in our previous papers.[3, 5]

Single crystal X-ray diffraction data were measured on a Bruker SMART APEX diffractometer equipped with an Apex II area detector and an Oxford Cryosystems 700 Series temperature controller. A crystal of BAs with dimensions $0.08 \times 0.07 \times 0.07$ mm was mounted by using Permatex High-Temp silicone glue. The crystal was measured between 100 K and 460 K with 20 K intervals. The ramping rate of the temperature was 2 K per minute. At each temperature, 660 frames were measured by using Mo K$\alpha$ radiation and a narrow-frame algorithm with scan widths of 0.5° in omega. The exposure time was 30 second/frame at 6 cm detector distance. The crystal was kept at the corresponding temperature for 10 minutes before each measurement. The lattice parameters were refined with the Bruker Apex-II program.

Neutron powder diffraction (NPD) data were collected on the NOMAD time-of-flight diffractometer of Oak Ridge National Laboratory's Spallation Neutron Source.[14] Data reduction and Rietveld refinements were performed using the ADDIE and GSAS software suites respectively.[15-17] As an internal temperature standard and to mitigate the strong neutron absorption of B, Si powder was mixed with BAs and BP samples in a mass ratio of 6:1 (Si:BAs(BP)). The final reported sample temperatures were determined by comparing the refined lattice parameter of the Si phase to previous reports on the thermal expansion of Si.[18, 19]

Both BAs and BP crystallize in the zinc-blende cubic structure with space group of $F\bar{4}3m$ (#216). The series of X-ray diffraction measurements carried out in the present study reconfirms the structure. The lattice parameters obtained from single crystal X-ray diffraction at different temperatures are shown in Table 1. The temperature range is from 100 K to 460 K with 20K interval for each point. The cell parameters were obtained by refining the 2θ values of about 120 reflections with 2θ varing between 15° and 60°. The standard deviation of the cell parameter determined at each temperature is on the order of $10^{-2}$ %. Qualitatively, the lattice parameter is seen to monotonically increase with increasing temperature as expected.

**Table I**. Lattice parameters of BAs and BP between 100K and 460K determined by X-ray single crystal diffraction

| T(K) | BAs | | BP | |
|---|---|---|---|---|
| | a(Å) | σ | a(Å) | σ |
| 100 | 4.7741 | 0.0015 | 4.5383 | 0.0008 |
| 120 | 4.7743 | 0.0014 | 4.5382 | 0.0008 |
| 140 | 4.7743 | 0.0014 | 4.5382 | 0.0008 |
| 160 | 4.7746 | 0.0015 | 4.5384 | 0.0009 |
| 180 | 4.7749 | 0.0014 | 4.5383 | 0.0008 |
| 200 | 4.7751 | 0.0015 | 4.5385 | 0.0008 |
| 220 | 4.7753 | 0.0015 | 4.5387 | 0.0008 |
| 240 | 4.7756 | 0.0015 | 4.5393 | 0.0008 |
| 260 | 4.7759 | 0.0015 | 4.5395 | 0.0008 |
| 280 | 4.7762 | 0.0014 | 4.5397 | 0.0008 |
| 300 | 4.7766 | 0.0014 | 4.5399 | 0.0008 |
| 320 | 4.7770 | 0.0014 | 4.5399 | 0.0008 |
| 340 | 4.7774 | 0.0015 | 4.5401 | 0.0008 |
| 360 | 4.7780 | 0.0015 | 4.5405 | 0.0008 |
| 380 | 4.7784 | 0.0015 | 4.5408 | 0.0009 |
| 400 | 4.7789 | 0.0015 | 4.5412 | 0.0008 |
| 420 | 4.7792 | 0.0015 | 4.5415 | 0.0008 |
| 440 | 4.7797 | 0.0015 | 4.5419 | 0.0008 |
| 460 | 4.7801 | 0.0015 | 4.5423 | 0.0008 |

Fig. 1 shows three representative neutron powder diffraction patterns for each BAs and BP sample with three selected temperature points being the lowest temperature (~300K), highest temperature (~1100K), and one temprature (~700K) point in between. Since the BAs starts to decompose to $B_{12}As_2$ ~1200K, the highest temperature used for the neutron diffraction is ~1150K. We note that the reported temperatures for the neutron powder diffraction data have a high degree of confidence due to the use of an internal Si standard during the measurements. The diffraction pattern can be modeled well using the cubic structure of BAs and BP with excellent Rp and wRp values. The detailed refinement data for BAs and BP with calibrated temperatures are shown in the Table 2 and Table 3 respectively for all the measured temperature range from 300K to 1150K. Both Rp and Rwp are small for BAs and BP at each temperature point, and the refined standard deveation of the cell parameter though GSAS is at each temperature is in the order of $10^{-4}$%. There is an overlap region in the X-ray and neutron diffraction data between 300K to 460K, where the refined cell parameters match well with each other between the two experiments. The room temperature lattice parameter is $a_{xrd}$ = 4.7766 ± 0.0014 Å and $a_{neutron}$ = 4.77659 ± 0.0003 Å for BAs, and room temperature lattice parameter is $a_{xrd}$ = 4.5399 ± 0.0014 Å and $a_{neutron}$ = 4.53991 ± 0.0003 Å for BP. As there is a negligible lattice parameter difference obtained from neutron and X-ray diffraction, we can combine both X-ray single crystal diffraction and neutron scattering data for a full range 100-1150K to determine the temperature dependence of lattice parameters and their subsequent CTE for both BAs and BP.

**Table II**. Lattice parameter and refined parameters of BAs between 300K and 1150K determined by neutron diffraction

| T(K) | a(Å) | σ | wR$_p$ | R$_p$ | $\chi^2$ |
|---|---|---|---|---|---|
| 299.8 | 4.77659 | 4.77E-04 | 0.0278 | 0.032 | 2.633 |
| 401.2 | 4.77841 | 4.85E-04 | 0.0277 | 0.0314 | 3.433 |
| 467.2 | 4.77988 | 4.87E-04 | 0.027 | 0.0316 | 3.216 |
| 543.7 | 4.78208 | 5.14E-04 | 0.0274 | 0.031 | 3.256 |
| 603.3 | 4.78322 | 5.35E-04 | 0.272 | 0.0315 | 3.235 |
| 667.6 | 4.7859 | 5.64E-04 | 0.0276 | 0.0312 | 3.3 |
| 743.3 | 4.78732 | 5.90E-04 | 0.0278 | 0.0313 | 3.399 |
| 779.9 | 4.78916 | 5.97E-04 | 0.0284 | 0.0316 | 3.483 |
| 848.4 | 4.79086 | 6.31E-04 | 0.0293 | 0.0314 | 3.699 |
| 919.2 | 4.79327 | 6.81E-04 | 0.0306 | 0.0329 | 3.559 |
| 1042.8 | 4.79712 | 7.36E-04 | 0.0319 | 0.0328 | 3.836 |
| 1097.5 | 4.79908 | 7.77E-04 | 0.0347 | 0.0359 | 2.851 |
| 1168.4 | 4.80154 | 8.21E-04 | 0.0343 | 0.0343 | 3.583 |

**Table III**. Lattice parameter and refined parameters of BP between 300K and 1150K determined by neutron diffraction

| T(K) | a(Å) | σ | wR$_p$ | R$_p$ | $\chi^2$ |
|---|---|---|---|---|---|
| 299.8 | 4.53991 | 3.04E-04 | 0.0352 | 0.0457 | 4.821 |
| 422.0 | 4.54219 | 3.04E-04 | 0.0314 | 0.0452 | 5.853 |
| 472.1 | 4.54323 | 3.04E-04 | 0.0313 | 0.0446 | 5.727 |
| 511.9 | 4.54427 | 3.04E-04 | 0.0311 | 0.0423 | 5.645 |
| 622.2 | 4.54617 | 3.04E-04 | 0.0321 | 0.045 | 5.87 |
| 664.9 | 4.54765 | 3.04E-04 | 0.0312 | 0.0423 | 5.539 |
| 720.4 | 4.54887 | 3.04E-04 | 0.0322 | 0.0436 | 5.788 |
| 788.4 | 4.55082 | 3.04E-04 | 0.0306 | 0.0421 | 5.211 |
| 826.7 | 4.55216 | 3.04E-04 | 0.03 | 0.0394 | 5.042 |
| 880.6 | 4.55387 | 3.04E-04 | 0.0288 | 0.0377 | 4.657 |
| 988.9 | 4.55621 | 3.04E-04 | 0.0314 | 0.0415 | 5.702 |
| 1015.8 | 4.55767 | 3.04E-04 | 0.0293 | 0.0371 | 4.925 |
| 1077.4 | 4.55977 | 3.04E-04 | 0.0293 | 0.0376 | 4.649 |
| 1110.4 | 4.56189 | 3.04E-04 | 0.0313 | 0.0406 | 3.1409 |
| 1148.2 | 4.56402 | 3.04E-04 | 0.0302 | 0.0401 | 5.023 |

The linear coefficient of thermal expansion can be expressed as

$$\alpha_L = \frac{1}{a}\frac{da(T)}{dT} \tag{1}$$

where a is the lattice parameter and d$a$(T)/dT is the differential lattice parameter over temperature. For the cubic system, the volumetric coefficient of thermal expansion can be simplified as $\alpha_V$=1/V dV/dT= = $1/a^3$ d$a^3$/dT = $3\alpha_L$. The temperature dependent of the lattice parameter from both X-ray (black square) and neutron scattering(red circle) with error bars for BAs sample is shown in Fig. 2a. The green line is the polynomial fit of the experiment data over the full temperature range, and the full data could be well fitted by third order polynomial curve, as shown below.

$$a_{BAs} = 4.77364 + 1.30\times10^{-6}T + 3.12\times10^{-8}T^2 - 1.03\times10^{-11}T^3 (\text{Å}) \tag{2}$$

By combination of Equation (1) and (2), we obtain the temperature dependent linear coefficient of thermal expansion for BAs shown in the Fig. 2b.

The temperature dependence of the thermal expansion is found analogous to the temperature dependent specific heat. For the commonly used Gruneison theory, the coefficient of thermal expansion can be expressed as

$$\alpha_L(T) = \frac{1}{3}\kappa\gamma C_v(T) \tag{3}$$

where κ is the harmonic compressibility, γ is the Gruneisen parameter, and $C_v$ is the phonon specific heat. If we make a simplification of Eq. (3) to treat harmonic compressibility and Gruneisen parameter as constants, then we will get $\alpha_L$ proportion to the phonon heat capacity. In the Debye model, phonon heat capacity can be expressed as

$$C_v = \frac{\partial U}{\partial T} = 9R(\frac{T}{T_D})^3 \int_0^{x_D} dx \frac{x^4 e^x}{(e^x - 1)^2} \tag{4}$$

where R is the gas constant, $T_D$ is the Debye temperature, and $x_D$=$T_D$/T. For temperatures higher than the Debye temperature of BAs (~700K), one should expect a general saturation effect in the CTE, in proportion to heat capacity behavior at high temperature.[20]

Since BP has a much higher Debye temperature (~1100K) than BAs (~700K)[1], one would expect a different temperature dependence of the CTE. Fig. 3 shows the polynomial fitting the BP lattice parameters. The third order polynomial fits the BP lattice parameter data reasonably well, however, there are some small discrepancy between the data and the fitting curve at the low temperature range. Both X-ray and heat capacity measurement have ruled out the possibility of phase transition at that temperature range. We further fit the data for higher order polynomial terms up to 6. As shown in the Fig. 3a, although the fourth order polynomial (yellow line) fits the data slightly better than third order polynomial fitting (green line) especially in the low temperature range, the resulting temperature dependent CTE exhibits a cubic graph curve shown as the inset of Fig. 3b. While the negative coefficient of thermal expansion determined from this fit at low temperatures might be expected in the BP (which need to be further validated), similar to the negative CTE behavior observed in the GaAs (< ~60K)[21] and Si (<120 K)[22] at low temperature. The increase of the CTE after the saturation beyond Debye temperature is apparently against the Gruneison theory we have described above and cannot be valid. Therefore, we conclude the third order polynomial best fits of our data, and is used for the following discussions.

The third order polynomial fitting for BP gives:

$$a_{BP} = 4.53698 + 6.75 \times 10^{-6}T + 1.29 \times 10^{-8}T^2 - 1.08 \times 10^{-11}T^3 (\text{Å}) \tag{5}$$

and by combination of Equation (1) and (5), we obtain the temperature dependent linear coefficient of thermal expansion for BP shown in the Fig. 3b.

Our obtained coefficient of thermal expansion at room temperature based on polynomial fitting is $3.6 \pm 0.15 \times 10^{-6}$/K for BAs and $3.2 \pm 0.2 \times 10^{-6}$/K for BP, respectively. These values are slightly higher than the theoretical predicted value $3.04 \times 10^{-6}$ K$^{-1}$ for BAs and $2.8 \times 10^{-6}$ K$^{-1}$ for BP, but indeed much better close to the CTE values of the commercially used semiconductor Si and GaAs in compassion with diamond and copper.

**Table 4**. Physical properties of some semiconducting and high thermal conductivity materials

| Compound | $a$(Å) | CTE($10^{-6}$/K) | $\kappa$(W/m/K) |
| --- | --- | --- | --- |
| Si | 5.431 | 2.63 | 130[23] |
| GaAs | 5.652 | 5.73 | 55[24] |
| GaN | $a$=3.190 $c$=5.189 | 3.17($c$), 5.59($a$) | 200[25] |
| Diamond | 3.567 | 0.8 | 2200[26] |
| SiC | 4.348 | 3.8 | 360[27] |
| BAs | 4.776 | 3.6 | 1300[2-4] |
| BP | 4.540 | 3.2 | 490[5] |
| Cu | 3.597 | 16.5 | 401[11] |

In Table 4 we list the lattice parameters, CTE and thermal conductivity for several comercially used semiconductor and potential heat spreaders. Compared to the other heat speader materials such as diamond and SiC, the BAs and BP have prevailing advantges with smaller lattice mismatch, and much better compatbility of CTE with Si and GaAs. Currently, for the diamond heat spreader application, in order to overcome the critical delamination and cracking issues caused by both large lattice mismatch and CTE imcompatbility, either multilayer configuration[28] or ceramic-reinforced metal matrix composite has been utilized to offer a better solution for tuning the both lattice mismatch and thermal expansion coefficient with the substrate, which unavoidably complicates the device architecture. Given much smaller lattice mismatch of BAs and BP with Si and GaAs, one now may expect to directly grow BAs and BP films on the semiconducting substrates, and use these high thermal conductivity materials as the heat spreader. In fact, eptaxial BP films have been grown succesfully on AlN and SiC[29, 30]. Their full potential in the semiconductor device application therefore looks very promising. It is worthwhile to mention that the temperature dependent lattice parameter behaviors of BAs/BP might be different with various substrates, and this needs to be carefully considered as well to minimize the thermal stress induced by cooling process after the film deposition.

The thermal expansion of the cubic BAs and BP was determined in the extended temperature range from 100K to 1150K, through a combination of X-ray single crystal diffraction and powder neutron diffraction. The obtained coefficient of thermal expansion at room temperature is $3.6 \pm 0.15 \times 10^{-6}$ /K for BAs and $3.2 \pm 0.2 \times 10^{-6}$ /K for BP, respectively, which have a much smaller thermal expansion mismatch to the common semiconductors as compared to other materials such as diamond and copper, The high thermal conductivity and compatible CTE with semiconductors, have placed BAs and BP as excellent candidates for the future heat spreader materials to solve increasing heat dispassion issues in the semiconducting device and power electronics.

Note added: Near the completion of our manuscript, we become aware of another independent study for BAs based on powder X-ray diffraction measurements from 300K to 773K, where a slightly higher $4.2 \pm 0.4 \times 10^{-6}$ /K CTE value is obtained.[31]

This work at University of Texas at Dallas is supported in part by Office of Naval Research (ONR) grant No. N00014-19-1-2061, and subcontract from ONR MURI grant N00014-16-1-2436. A portion of this work used resources at the Spallation Neutron Source, a DOE Office of Science User Facility operated by the Oak Ridge National Laboratory.

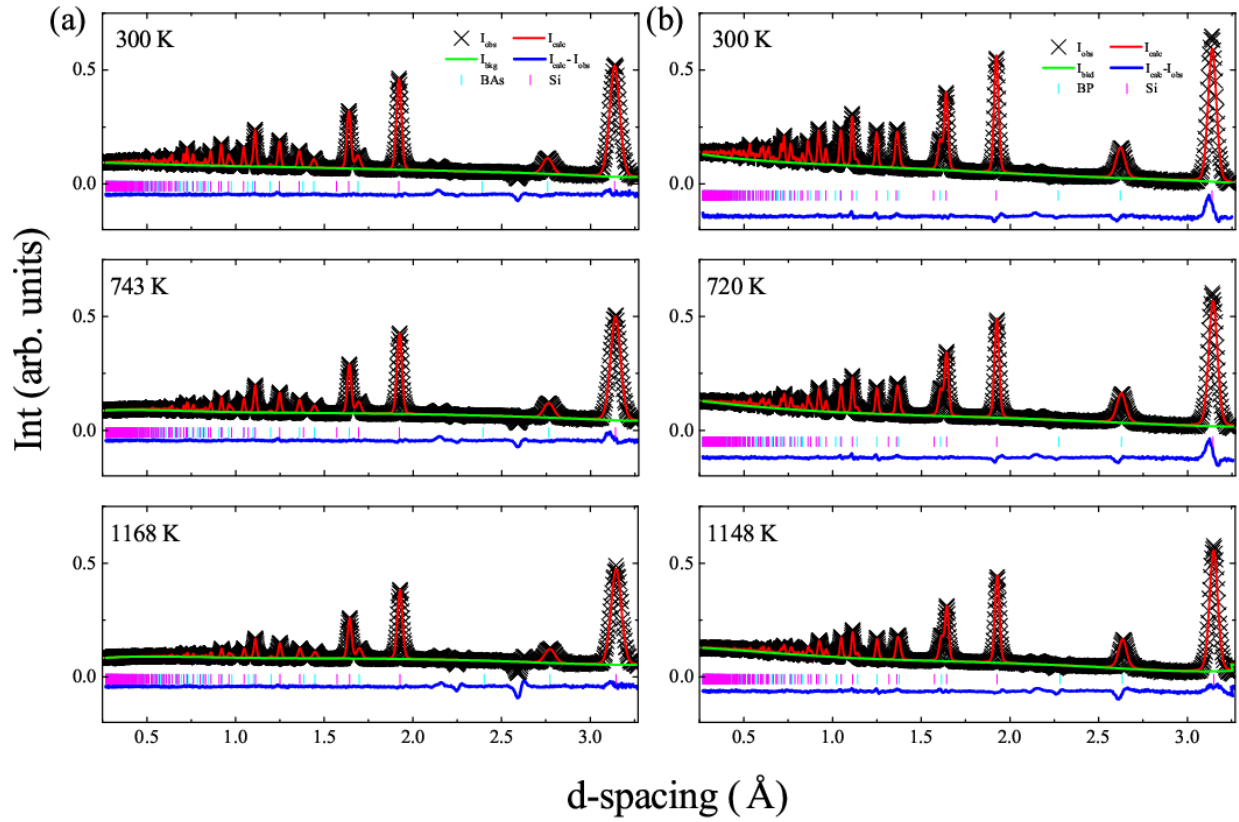

FIG. 1 Neutron scattering data for (a) BAs and (b) BP at different temperatures together with the best fit models from Rietveld refinements. The data was reduced using ADDIE and refinements were performed using GSAS. We note that the slight dips observed at ~2.2 and 2.55 Å are due to artifacts of normalization arising from the strong neutron absoption of B.

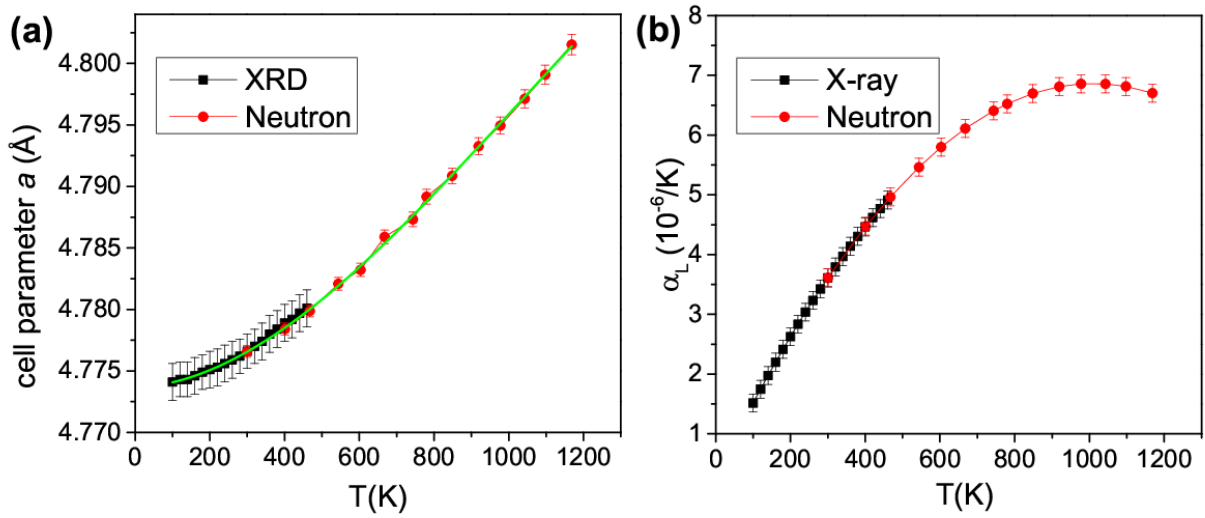

*FIG. 2 (a) Combined full range lattice parameter for BAs, the green line is the fitted curve using third order polynomial. (b) Linear CTE data of BAs derived from lattice parameter.*

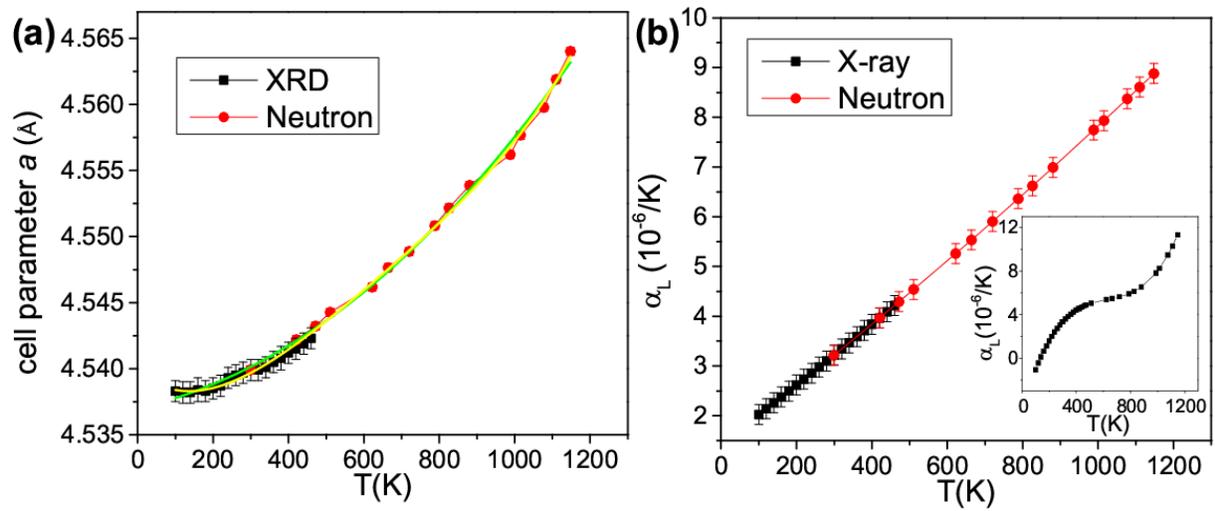

*FIG. 3 Combined full range lattice parameter for BP, the green and yellow lines are the fitted curves using third and fourth order polynomial respectively. (b) Linear CTE data of BP derived from third order polynomial fitting, insert is from fourth order polynomial fitting results.*